\documentclass[12pt]{article}
\usepackage{graphicx}
\usepackage{lscape}
\usepackage{citesort}
\usepackage{amssymb}
\usepackage{appendix}
\usepackage{multirow}

\begin{document}
\def\qq{\langle \bar q q \rangle}
\def\uu{\langle \bar u u \rangle}
\def\dd{\langle \bar d d \rangle}
\def\sp{\langle \bar s s \rangle}
\def\GG{\langle g_s^2 G^2 \rangle}
\def\Tr{\mbox{Tr}}
\def\figt#1#2#3{
        \begin{figure}
        $\left. \right.$
        \vspace*{-2cm}
        \begin{center}
        \includegraphics[width=10cm]{#1}
        \end{center}
        \vspace*{-0.2cm}
        \caption{#3}
        \label{#2}
        \end{figure}
	}
	
\def\figb#1#2#3{
        \begin{figure}
        $\left. \right.$
        \vspace*{-1cm}
        \begin{center}
        \includegraphics[width=10cm]{#1}
        \end{center}
        \vspace*{-0.2cm}
        \caption{#3}
        \label{#2}
        \end{figure}
                }

\def\ds{\displaystyle}
\def\beq{\begin{equation}}
\def\eeq{\end{equation}}
\def\bea{\begin{eqnarray}}
\def\eea{\end{eqnarray}}
\def\beeq{\begin{eqnarray}}
\def\eeeq{\end{eqnarray}}
\def\ve{\vert}
\def\vel{\left|}
\def\ver{\right|}
\def\nnb{\nonumber}
\def\ga{\left(}
\def\dr{\right)}
\def\aga{\left\{}
\def\adr{\right\}}
\def\lla{\left<}
\def\rra{\right>}
\def\rar{\rightarrow}
\def\lrar{\leftrightarrow}  
\def\nnb{\nonumber}
\def\la{\langle}
\def\ra{\rangle}
\def\ba{\begin{array}}
\def\ea{\end{array}}
\def\tr{\mbox{Tr}}
\def\ssp{{\Sigma^{*+}}}
\def\sso{{\Sigma^{*0}}}
\def\ssm{{\Sigma^{*-}}}
\def\xis0{{\Xi^{*0}}}
\def\xism{{\Xi^{*-}}}
\def\qs{\la \bar s s \ra}
\def\qu{\la \bar u u \ra}
\def\qd{\la \bar d d \ra}
\def\qq{\la \bar q q \ra}
\def\gGgG{\la g^2 G^2 \ra}
\def\q{\gamma_5 \not\!q}
\def\x{\gamma_5 \not\!x}
\def\g5{\gamma_5}
\def\sb{S_Q^{cf}}
\def\sd{S_d^{be}}
\def\su{S_u^{ad}}
\def\sbp{{S}_Q^{'cf}}
\def\sdp{{S}_d^{'be}}
\def\sup{{S}_u^{'ad}}
\def\ssp{{S}_s^{'??}}

\def\sig{\sigma_{\mu \nu} \gamma_5 p^\mu q^\nu}
\def\fo{f_0(\frac{s_0}{M^2})}
\def\ffi{f_1(\frac{s_0}{M^2})}
\def\fii{f_2(\frac{s_0}{M^2})}
\def\O{{\cal O}}
\def\sl{{\Sigma^0 \Lambda}}
\def\es{\!\!\! &=& \!\!\!}
\def\ap{\!\!\! &\approx& \!\!\!}
\def\ar{&+& \!\!\!}
\def\ek{&-& \!\!\!}
\def\kek{\!\!\!&-& \!\!\!}
\def\cp{&\times& \!\!\!}
\def\se{\!\!\! &\simeq& \!\!\!}
\def\eqv{&\equiv& \!\!\!}
\def\kpm{&\pm& \!\!\!}
\def\kmp{&\mp& \!\!\!}
\def\mcdot{\!\cdot\!}
\def\erar{&\rightarrow&}


\def\simlt{\stackrel{<}{{}_\sim}}
\def\simgt{\stackrel{>}{{}_\sim}}


\title{
         {\Large
                 {\bf
The masses and residues of doubly heavy spin--3/2 baryons
                 }
         }
      }

\author{\vspace{1cm}\\
{\small T. M. Aliev \thanks {e-mail:
taliev@metu.edu.tr}~\footnote{permanent address:Institute of
Physics,Baku,Azerbaijan}\,\,, K. Azizi \thanks {e-mail:
kazizi@dogus.edu.tr}\,\,, M. Savc{\i} \thanks
{e-mail: savci@metu.edu.tr}} \\
{\small Physics Department, Middle East Technical University,
06531 Ankara, Turkey }\\
{\small$^\ddag$ Physics Department,  Faculty of Arts and Sciences,
Do\u gu\c s University,} \\
{\small Ac{\i}badem-Kad{\i}k\"oy,  34722 Istanbul, Turkey}}

\date{}

\begin{titlepage}
\maketitle
\thispagestyle{empty}

\begin{abstract}

The masses and residues of the spin--3/2 doubly heavy baryons are calculated
within the QCD sum rules method. A comparison of our predictions with those existing in the literature is also made.

\end{abstract}

~~~PACS numbers: 11.55.Hx, 14.20.--c, 14.20.Mr

\end{titlepage}

\section{Introduction}

During the last few years, there has been substantial experimental progress
on the spectroscopy of the light and heavy baryons. Among the discovered many
new states, some are described by the quark model, but the quark content of
some others is under debate. Many new states containing a single heavy quark are
experimentally observed. Practically, all baryons containing a single charm
quark that predicted by the quark model  have already been established.
The heavy spin--1/2, $\Lambda_b$, $\Sigma_b$, $\Xi_b$ and $\Omega_b$ baryons as well as the
spin--3/2, $\Sigma_b^\ast$ baryon containing a single $b$--quark are also
experimentally discovered (for a review, see for example \cite{Rdhbtt01}).
Recently, the CMS Collaboration announced the observation of the spin--3/2, 
$\Xi_b^\ast$ baryon containing also a single b--quark with a mass of $5945~MeV$
\cite{Rdhbtt02}. 

The quark model also predicts hadrons with two or three heavy quarks.
The experimental progress on the heavy hadron spectroscopy has stimulated the researches on
the doubly heavy hadron physics. Up to now, only one heavy baryon with two
charm quarks, namely the $\Xi_{cc}^+$ baryon has been observed in experiments
\cite{Rdhbtt03,Rdhbtt04,Rdhbtt05} conducted by the SELEX Collaboration. Researches
are waiting  for considerable experimental progress on observations of the doubly heavy baryons and their properties at LHC--b.

One of the main characteristic parameters of the  doubly heavy baryons is their mass.
The  masses of these baryons are tried to be estimated in different frameworks such as quark
model \cite{Rdhbtt06,Rdhbtt07} and  MIT bag model \cite{Rdhbtt08}. The masses of
the spin--1/2 doubly heavy baryons are also calculated within QCD sum rules method
\cite{Rdhbtt09} in \cite{Rdhbtt10,Rdhbtt11,Rdhbtt12,Rdhbtt13}, and the  masses of
the spin--3/2 doubly heavy baryons are studied within the same framework in
\cite{Rdhbtt10,Rdhbtt11} and \cite{Rdhbtt14}. However, one can easily see
that the analytical expressions presented in these three works are different, and
therefore there appears a necessity for a more accurate study of the masses
and residues of the doubly heavy baryons within the QCD sum rules method.

The outline of the paper is as follows. In section 2, we obtain the sum rules for the masses and residues of the  doubly heavy baryons. In section 3, we present the numerical
analysis of the sum rules and discuss the results. We also compare the obtained results with those predicted via other nonperturbative approaches in this section. 

\section{Mass sum rules for the doubly heavy spin--3/2 baryons}

The QCD sum rules for the doubly heavy spin--3/2 baryons are obtained by
considering the two--point correlator,
\bea
\label{edhbtt01}
\Pi_{\mu\nu} (q) = i \int d^4x e^{iqx} \lla 0 \vel {\cal T} \{ \eta_\mu (x)
\bar{\eta}_\nu (0) \} \ver 0 \rra ~,
\eea
where ${\cal T}$ is the time ordering operator,  $q$ is the four--momentum of the doubly heavy baryon and $\eta_\mu$
is its interpolating current.
Few words about the choices of the interpolating current for the spin--3/2
doubly heavy baryons are in order. The general structure of the interpolating
current should contain the following terms: 
$\varepsilon^{abc} ( Q^{aT} C \Gamma Q^{'b} ) \widetilde{\Gamma} q^c$,
$\varepsilon^{abc} (q^{aT} C \Gamma Q^b ) \widetilde{\Gamma} Q^{'c}$,
and
$\varepsilon^{abc} (q^{aT} C \Gamma Q^{'b}) \widetilde{\Gamma} Q^c$,
where $T$ is the transposition, $C$ is the charge conjugation operator, $\Gamma$ and $\widetilde{\Gamma}$ are Dirac matrices;
and $a,~b$ and $c$ are the color indices. Since we are interested in the doubly heavy
baryons with spin--3/2, each diquark in the above--presented forms should
obviously have spin 1. As far as the first term is concerned, since the
diquark has spin 1, it should be symmetric with respect to the  
$Q \lrar Q^{'}$ exchange. This implies that $\Gamma$ is to be replaced by
$\gamma_\mu$ or $\sigma_{\mu\nu}$. The remaining two terms should
also exhibit this symmetry property. Hence, these two terms should
have the following form:
\bea
\label{nolabel01}
\varepsilon^{abc} \Big[(q^{aT} C \Gamma Q^b ) \widetilde{\Gamma} Q^{'c} +
(q^{aT} C \Gamma Q^{'b} ) \widetilde{\Gamma} Q^c \Big]~, \nnb
\eea
where $\Gamma = \gamma_\mu$ or $\sigma_{\mu\nu}$.

As a result, the two possible forms of the interpolating current for the 
double heavy baryons can be written as:
\bea
\label{nolabel02}
&&N_1 \varepsilon^{abc} \Big\{ ( Q^{aT} C \gamma_\mu Q^{'b} )
\widetilde{\Gamma}_1 q^c + (q^{aT} C \gamma_\mu Q^b ) \widetilde{\Gamma}_1
Q^{'c} + (q^{aT} C \gamma_\mu Q^{'b} ) \widetilde{\Gamma}_1 Q^c \Big\}~,~\\
\mbox{\rm or}, \nnb \\
&&N_2 \varepsilon^{abc} \Big\{ ( Q^{aT} C \sigma_{\mu\nu} Q^{'b} )      
\widetilde{\Gamma}_2 q^c + (q^{aT} C \sigma_{\mu\nu} Q^b ) \widetilde{\Gamma}_2  
Q^{'c} + (q^{aT} C \sigma_{\mu\nu} Q^{'b}) \widetilde{\Gamma}_2 Q^c
\Big\}~,\nnb
\eea
where $N_1$ and $N_2$ are the normalization factors. The values of
$\widetilde{\Gamma}_1$ and $\widetilde{\Gamma}_2$ are determined through a
consideration involving Lorentz structure and parity. Since the
above--mentioned forms must both be Lorentz vectors, then
$\widetilde{\Gamma}_1=1$ or $\gamma_5$, and
$\widetilde{\Gamma}_2=\gamma_\nu$ or $\gamma_5\gamma_\nu$. Furthermore, the parity
consideration leads to the results $\widetilde{\Gamma}_1=1$ and
$\widetilde{\Gamma}_2=\gamma_\nu$.
Thus, as a result of the above discussion, we have two possible interpolating
currents for the spin--3/2 doubly heavy baryons,
\bea
\label{nolabel03}
\eta_{1\mu} \es N_1 \varepsilon^{abc} \Big\{ ( Q^{aT} C \gamma_\mu Q^{'b} ) q^c +
(q^{aT} C \gamma_\mu Q^b ) Q^{'c} + (q^{aT} C \gamma_\mu Q^{'b} ) Q^c \Big\}~, \nnb \\  
\eta_{2\mu} \es N_2 \varepsilon^{abc} \Big\{ 
( Q^{aT} C \sigma_{\mu\nu} Q^{'b} )\gamma_\nu q^c +
(q^{aT} C \sigma_{\mu\nu} Q^b ) \gamma_\nu  Q^{'c} + (q^{aT} C \sigma_{\mu\nu}
Q^{'b} ) \gamma_\nu Q^c
\Big\}~.\nnb
\eea
Moreover, if we formally assume that all quarks are heavy (light) and
$Q^{'}=Q$, only the  $\eta_{1\mu}$ survives  similar to the $\Delta^{++}$
current.
For this reason, in the present work, we consider the following current as the
interpolating current for doubly heavy baryons with spin--3/2:
\bea
\label{edhbtt02}
\eta_\mu = {1\over \sqrt{3}} \epsilon^{abc} \Big\{
(q^{aT} C \gamma_\mu Q^b) Q^{\prime c} +
(q^{aT} C \gamma_\mu Q^{\prime b}) Q^c +
(Q^{aT} C \gamma_\mu Q^{\prime b}) q^c \Big\}~,
\eea
where  $q$ is the light; and $Q$ and $Q^\prime$
are the two  heavy quarks, respectively. We present the quark content of the doubly
heavy baryons in Table 1.

\begin{table}[h]

\renewcommand{\arraystretch}{1.3}
\addtolength{\arraycolsep}{-0.5pt}
\small
$$
\begin{array}{|l|c|c|c|c|}
\hline \hline   
  \mbox{baryon}           & \mbox{Light quark $q$} & \mbox{Heavy quark $Q$}  & \mbox{Heavy quark $Q^\prime$}\\  \hline
 \Xi_{Q Q}^\ast           & u~\mbox{or}~d          & b~\mbox{or}~c           & b~\mbox{or}~c               \\
 \Omega_{Q Q}^\ast        & s                      & b~\mbox{or}~c           &b~\mbox{or}~c                \\
 \Xi_{Q Q^\prime}^\ast    & u~\mbox{or}~d          & b                       & c                            \\
 \Omega_{Q Q^\prime}^\ast & s                      & b                       &             c                \\
\hline \hline
\end{array}
$$
\caption{The quark content of the spin--3/2 doubly heavy
baryons.}
\renewcommand{\arraystretch}{1}
\addtolength{\arraycolsep}{-1.0pt}
\end{table}

As is well known, in the QCD sum rules approach, the correlation function is
calculated in two different manners:
\begin{itemize}

\item In terms of quarks and gluons using the operator product expansion
(OPE), which contains perturbative and nonperturbative condensate
contributions,

\item In terms of hadrons (physical part).

\end{itemize}

Equating these two representations and performing Borel transformation with
respect to the baryon momentum square, which suppresses the higher states and
continuum contributions, we obtain the sum rules. Here we would like to make the
following cautionary note. The interpolating current $\eta_\mu$ of the
doubly heavy baryons can interact not only with the positive parity spin--3/2
baryons, but also with the negative parity spin--3/2 baryons, as well as
with spin--1/2 baryons with both parities, and surely these unwanted
contributions must be eliminated.

The matrix element of the interpolating current $\eta_\mu$ sandwiched 
between the vacuum and the single baryon states is determined in the
following way:
\bea
\label{edhbtt03}
\lla 0 \vel \eta_\mu \ver B_{(3/2)^+}(q) \rra \es \lambda_{(3/2)^+}
u_\mu (q)~, \nnb \\ 
\lla 0 \vel \eta_\mu \ver B_{(3/2)^-}(q) \rra \es \lambda_{(3/2)^-}
\gamma_5 u_\mu (q)~, \nnb \\
\lla 0 \vel \eta_\mu \ver B_{(1/2)^+}(q) \rra \es \lambda_{(1/2)^+}
\Bigg( {4 q_\mu \over m} + \gamma_\mu \Bigg) \gamma_5 u(q), \nnb \\
\lla 0 \vel \eta_\mu \ver B_{(1/2)^-}(q) \rra \es \lambda_{(1/2)^-}
\Bigg( {- 4 q_\mu \over m} + \gamma_\mu \Bigg) u(q)~,
\eea
where $u_\mu$ is the Rarita--Schwinger spinor, and $\lambda_i$ are the
residues.

Now, we can proceed calculating the
physical part of the correlator given by Eq. (\ref{edhbtt01}). Saturating
this correlator by the ground state baryons we get,
\bea
\label{edhbtt04}
\Pi_{\mu\nu} = {\lla 0 \vel \eta_\mu \ver B(q) \rra \lla B(q) \vel
\bar{\eta}_\nu \ver 0 \rra\over q^2-m_B^2} + \cdots~,
\eea
where dots represent the higher states and continuum contributions.

Using Eqs. (\ref{edhbtt03}) and (\ref{edhbtt04}) and performing summation
over spins of the Rarita--Schwinger spinor which is given by the relation,
\bea
\label{edhbtt05}
\sum u_\mu(q,s) \bar{u}_\nu (q,s) = (\rlap/{q} + m_B) \Bigg( g_{\mu\nu} - {1
\over 3} \gamma_\mu \gamma_\nu - {2 q_\mu q_\nu \over 3 m_B^2} + {q_\mu
\gamma_\nu - q_\nu \gamma_\mu \over 3 m_B} \Bigg)~,
\eea
we obtain the following expression for the physical part of the correlation function:
\bea
\label{edhbtt06}
\Pi_{\mu\nu}(q) \es
{\lambda^2_{(3/2)^+} \over m_{(3/2)^+}^2 - q^2} (\rlap/{q} + m_{(3/2)^+})
\Bigg( g_{\mu\nu} - {1\over 3} \gamma_\mu \gamma_\nu -
{2 q_\mu q_\nu \over m_{(3/2)^+}^2} + {q_\mu\gamma_\nu - q_\nu\gamma_\mu
\over 3 m_{(3/2)^+}} \Bigg)~, \nnb \\
\ek {\lambda^2_{(3/2)^-} \over m_{(3/2)^-}^2 - q^2} \gamma_5 (\rlap/{q} + m_{(3/2)^-}) 
\Bigg( g_{\mu\nu} - {1\over 3} \gamma_\mu \gamma_\nu -
{2 q_\mu q_\nu \over m_{(3/2)^-}^2} + {q_\mu\gamma_\nu - q_\nu\gamma_\mu
\over 3 m_{(3/2)^-}} \Bigg) \gamma_5~, \nnb \\ 
\ek {\lambda^2_{(1/2)^+} \over m_{(1/2)^+}^2 - q^2} \Bigg( {4 q_\mu \over
m_{(1/2)^+}} +\gamma_\mu \Bigg) \gamma_5 (\rlap/{q} + m_{(1/2)^+}) 
\Bigg( {4 q_\nu \over m_{(1/2)^+}} +\gamma_\nu \Bigg) \gamma_5 \nnb \\
\ar {\lambda^2_{(1/2)^-} \over m_{(1/2)^-}^2 - q^2} \Bigg( - {4 q_\mu \over
m_{(1/2)^-}} +\gamma_\mu \Bigg) (\rlap/{q} + m_{(1/2)^-}) 
\Bigg( {- 4 q_\nu \over m_{(1/2)^-}} +\gamma_\nu \Bigg)~.
\eea

It follows from this expression that only the structures $\rlap/{q}
g_{\mu\nu}$ and $g_{\mu\nu}$ couple to the spin--3/2 baryons, which we shall consider in
further discussion. Therefore, for the physical part of the correlator we get,
\bea
\label{edhbtt07} 
\Pi_{\mu\nu}(q) \es {\lambda^2_{(3/2)^+} \over m_{(3/2)^+}^2 - q^2} (\rlap/{q}
+ m_{(3/2)^+}) g_{\mu\nu} +
{\lambda^2_{(3/2)^-} \over m_{(3/2)^-}^2 - q^2} (\rlap/{q} -
m_{(3/2)^-})  g_{\mu\nu} + \cdots
\eea

We now return our attention to the calculation of the correlator from the
QCD side. This calculation is carried out in deep Euclidean region using the
OPE. After some calculations, we obtain expression of the correlator for the baryons containing two different heavy quarks in terms
of light and heavy quarks propagators as follows:
\bea
\label{edhbtt08}
\Pi_{\mu\nu} (q) \es {1\over 3} \epsilon^{abc} \epsilon^{a^\prime b^\prime
c^\prime} \Big\{
-S_Q^{c b^\prime} \gamma_\nu \widetilde{S}_{Q^\prime}^{a a^\prime} \gamma_\mu
S_q^{b c^\prime} - S_Q^{c a^\prime} \gamma_\nu \widetilde{S}_{q}^{b b^\prime}
\gamma_\mu S_{Q^\prime}^{a c^\prime} -
S_{Q^\prime}^{c a^\prime} \gamma_\nu \widetilde{S}_{Q}^{b b^\prime} \gamma_\mu
S_{q}^{a c^\prime} \nnb \\
\ek S_{Q^\prime}^{c b^\prime} \gamma_\nu \widetilde{S}_{q}^{a a^\prime}
\gamma_\mu S_{Q}^{b c^\prime} - S_q^{c a^\prime} \gamma_\nu
\widetilde{S}_{Q^\prime}^{b b^\prime} \gamma_\mu S_Q^{a c^\prime} -
S_q^{c b^\prime} \gamma_\nu
\widetilde{S}_{Q}^{a a^\prime} \gamma_\mu S_{Q^\prime}^{b c^\prime} \nnb \\
\ek S_{Q^\prime}^{c c^\prime}  \mbox{Tr}\Big[S_{Q}^{b a^\prime} \gamma_\nu
\widetilde{S}_{q}^{a b^\prime} \gamma_\mu \Big] -
S_{q}^{c c^\prime}  \mbox{Tr}\Big[S_{Q^\prime}^{b a^\prime} \gamma_\nu
\widetilde{S}_{Q}^{a b^\prime} \gamma_\mu \Big] -
S_{Q}^{c c^\prime}  \mbox{Tr}\Big[S_{q}^{b a^\prime} \gamma_\nu
\widetilde{S}_{Q^\prime}^{a b^\prime} \gamma_\mu \Big] \Big\}~,
\eea
where $\widetilde{S} = C S^T C$.

It follows from Eq. (\ref{edhbtt08}) that, in order to calculate the correlator
from the QCD side, the expressions of the heavy and light quarks propagators
are needed. Their expressions in the coordinate representation are given as,
\bea         
\label{edhbtt09}
S_q(x) \es i {\rlap/{x} \over 2 \pi^2 x^4} - {m_q \over 4 \pi^2 x^2} - {\qq
\over 12} \Bigg( 1- i {m_q \over 4} \rlap/{x} \Bigg) - {x^2\over 192} m_0^2
\qq \Bigg( 1- i {m_q \over 6} \rlap/{x} \Bigg)~, \nnb \\
S_Q (x) \es {m_Q^2 \over 4 \pi^2} {K_1 (m_Q \sqrt{-x^2}) \over \sqrt{-x^2}
} - {m_Q^2 \rlap/{x} \over 4 \pi^2 x^2} K_2 (m_Q \sqrt{-x^2})~,
\eea
where $K_1$ and $K_2$ are the modified Bessel functions of the
second kind.

It should be noted here that the propagators contain also pieces proportional to
the gluon field strength tensor. While we perform numerical analysis
with these terms, we see that their contributions are very small, and for
this reason we do not present them in Eq. (\ref{edhbtt09}).
In calculating the correlator from QCD side with strange quark, we take its
mass in linear order.

The correlation function for the structure $\rlap/{q} g_{\mu\nu}$ or
$g_{\mu\nu}$ in QCD side can be written in terms of the dispersion relation as,
\bea
\label{edhbtt10}
\Pi_i (q^2) = \int_{(m_Q + m_{Q^\prime})^2}^\infty ds
{\rho_i \over s - q^2}~,
\eea
where $i=1(2)$ corresponds to the structure $\rlap/{q} g_{\mu\nu}$
($g_{\mu\nu}$). The spectral density $\rho_i $ in Eq. (\ref{edhbtt10})
is given by the imaginary part of the correlator,
\bea
\label{nolabel04}
\rho_i(s) = {1\over \pi} \mbox{Im}\Pi_i(s)~. \nnb
\eea

After tedious calculations, for the spectral densities we get,
\bea
\label{edhbtt11}
\rho_1 (s) \es {1\over 32 \pi^4} 
\int_{\alpha_{min}}^{\alpha_{max}} d\alpha
\int_{\beta_{min}}^{\beta_{max}} d\beta
\Big\{ \mu \Big[ 3 \alpha \beta (\alpha+\beta) \mu - 4 ( 1-\alpha-\beta)
m_{Q^\prime} m_Q \nnb \\
\ek 4 m_q (\alpha m_{Q^\prime} + \beta m_Q) \Big] \Big\} -
{\qq \over 24 \pi^2} \int_{\alpha_{min}}^{\alpha_{max}} d\alpha \Big[
(1-\alpha) (3 \alpha m_q - 4 m_Q ) - 4 \alpha m_{Q^\prime} \Big]~ \\ \nnb \\
\label{edhbtt12}
\rho_2 (s) \es {1\over 16 \pi^4}
\int_{\alpha_{min}}^{\alpha_{max}} d\alpha              
\int_{\beta_{min}}^{\beta_{max}} d\beta              
\Big\{ \mu \Big[ (\alpha+\beta) \mu (\alpha m_{Q^\prime} + \beta m_Q) -
m_q (\alpha \beta \mu + 3 m_Q m_{Q^\prime}) \Big] \Big\} \nnb \\
\ar {\qq \over 12 \pi^2} \int_{\alpha_{min}}^{\alpha_{max}} d\alpha \Big\{
(1-\alpha) \Big[2 \alpha (m_0^2 + \mu_1 -s ) - m_Q m_q \Big]+ (\alpha m_q + 3
m_Q) m_{Q^\prime} \Big\}~,
\eea 
where,
\bea
\label{nolabel05}
\mu \es {m_Q^2 \over \alpha} +  {m_{Q^\prime}^2 \over \beta} -s~, \nnb \\
\mu_1 \es \mu (\beta \to 1-\alpha)~, \nnb \\
\beta_{min} \es {\alpha m_{Q^\prime}^2 \over s\alpha - m_Q^2}~, \nnb \\
\beta_{max} \es 1 - \alpha~, \nnb \\
\alpha_{min} \es {1\over 2s} \Big[ s + m_Q^2 - m_{Q^\prime}^2 -
\sqrt{(s+m_Q^2 - m_{Q^\prime}^2)^2 - 4 m_Q^2 s}~, \nnb \\
\alpha_{max} \es {1\over 2s} \Big[ s + m_Q^2 - m_{Q^\prime}^2 +
\sqrt{(s+m_Q^2 - m_{Q^\prime}^2)^2 - 4 m_Q^2 s}~. \nnb
\eea

We can now compare our results on the spectral densities with the ones  
presented for instance in \cite{Rdhbtt10} and \cite{Rdhbtt11}. As far as the spectral
density $\rho_1(s)$ is concerned, we have the factor $3 (\alpha+\beta)$,
which is absent in \cite{Rdhbtt11}. The quark condensate in our case 
contains the term proportional to ${1\over 8}\alpha (1-\alpha) m_q$, while the
corresponding term in \cite{Rdhbtt11} is ${17 \over 48} \alpha (1-\alpha)
m_q$. The differences for the spectral density $\rho_2(s)$ can be summarized
as follows. The perturbative term proportional to $m_Q(m_{Q^\prime})$ (note
that $m_Q(m_{Q^\prime})$ in our work correspond to $m_{Q^\prime}(m_Q)$ in
\cite{Rdhbtt11}) contains the factor $(\alpha+\beta)$ which is
absent in \cite{Rdhbtt11}. In the perturbative part of $\rho_2(s)$ we have also
\bea
\label{nolabel06}
-{1 \over 16 \pi^4} m_q \int {d\alpha \over \alpha} \int {d\beta \over
\beta} (m_Q^2 \beta + m_{Q^\prime}^2 \alpha - s \alpha \beta)^2~, \nnb
\eea
which is again absent in \cite{Rdhbtt11}. When we compare the quark
condensate terms we have $4 \alpha(1-\alpha) s$ which is different from their
term reading $3 \alpha(1-\alpha) s$.

For the $\rlap/{q} g_{\mu\nu}$ structure, our results on perturbative part and
quark condensate terms without the strange quark mass agree with the ones
given in \cite{Rdhbtt10}, but the terms proportional to $m_s$ (which is
calculated in \cite{Rdhbtt15}) and the results for the $d=5$ operators are
different compared to those given in \cite{Rdhbtt10} and \cite{Rdhbtt15}.

Equating the coefficient of the structure $\rlap/{q} g_{\mu\nu}(g_{\mu\nu})$ in Eq.
(\ref{edhbtt07}) to Eq. (\ref{edhbtt10}) for $\Pi_1(\Pi_2)$, and performing
Borel transformation with respect to $Q^2=-q^2$, we get the following sum rules for
the masses and residues:
\bea
\label{edhbtt13}
\lambda_{(3/2)^+}^2 e^{-m_{(3/2)^+}^2/M^2} +
\lambda_{(3/2)^-}^2 e^{-m_{(3/2)^-}^2/M^2} \es 
\int_{(m_Q+m_Q^\prime)^2}^{s_0} ds \rho_1(s)e^{-s/M^2}~, \\
\label{edhbtt14}
\lambda_{(3/2)^+}^2 m_{(3/2)^+} e^{-m_{(3/2)^+}^2/M^2} -
\lambda_{(3/2)^-}^2 m_{(3/2)^-} e^{-m_{(3/2)^-}^2/M^2} \es 
\int_{(m_Q+m_Q^\prime)^2}^{s_0} ds \rho_2(s)
e^{-s/M^2}~,
\eea
in which the quark--hadron duality is used, and the contributions
of the higher states and continuum are modeled as the perturbative ones
starting from some threshold $s_0$. 

It follows from these sum rules that the negative parity spin--3/2 baryons
``contaminates" the sum rules. In order to eliminate
contributions of the $(3/2)^-$ baryons we multiply Eq. (\ref{edhbtt13}) with
$m_{(3/2)^-}$ and add it to Eq. (\ref{edhbtt14}), as a result of which we
get the following sum rule:
\bea
\label{edhbtt15}
\lambda_{(3/2)^+}^2 (m_{(3/2)^+}+m_{(3/2)^-}) e^{-m_{(3/2)^+}^2/M^2} =
\int_{(m_Q+m_Q^\prime)^2}^{s_0} ds \Big[m_{(3/2)^-} \rho_1(s) +
\rho_2(s) \Big] e^{-s/M^2}~.
\eea

It should be remembered that this approach is also used in estimating the coupling
constant of the  pseudoscalar mesons with heavy baryons containing single heavy
quark in \cite{Rdhbtt16}.

\section{Numerical analysis}

In this section, we present our numerical results on the masses and residues of
the spin--$\frac{3}{2}^+$ doubly heavy baryons. For the quark masses, we use their
$\overline{\mbox{MS}}$ values: $\bar{m}_c(\bar{m}_c) = (1.28 \pm 0.03)~GeV$,
$\bar{m}_b(\bar{m}_b) = (4.16 \pm 0.03)~GeV$ (see for example \cite{Rdhbtt16}),
and $m_s(2 ~GeV) = (102\pm8)~MeV$ \cite{Rdhbtt17}. The values of the quark
condensates are taken as $\langle \bar uu\rangle (1~GeV) = \langle \bar dd\rangle (1~GeV) =- (246_{-19}^{+28}~MeV)^3$
\cite{Rdhbtt18}, $\langle \bar s s\rangle=0.8\langle \bar u u\rangle$ and $m_0^2=(0.8\pm0.2)~GeV^2$. The masses of the negative parity doubly heavy baryons 
are taken from \cite{Rdhbtt19}, in which the QCD sum rules have been used
in calculating them. These masses are calculated to have the following
values: $m_{\Xi^*_{cc}}(\frac{3}{2}^-) = (3.80 \pm 0.18)~GeV$,  $m_{\Omega^*_{cc}}(\frac{3}{2}^-) = (3.96 \pm
0.16)~GeV$,  $m_{\Xi^*_{bb}}(\frac{3}{2}^-) = (10.43 \pm 0.15)~GeV$ and  $m_{\Omega^*_{bb}}(\frac{3}{2}^-) =
(10.57 \pm 0.15)~GeV$.

It should be noted that the masses of the negative parity spin--3/2 baryons
$\Xi^*_{bc}(\frac{3}{2}^-)$ and $\Omega^*_{bc}(\frac{3}{2}^-)$ are not estimated in \cite{Rdhbtt19}. We
observe that the mass difference of the positive and negative parity baryons
with two identical heavy quarks is about $200~MeV$, and estimate that
similar difference could have existed for $\Xi^*_{bc}$ and $\Omega^*_{bc}$
type baryons. So, we take the mass of these negative parity baryons as,
$m_{\Xi^*_{bc}}(\frac{3}{2}^-) = 7.4~GeV$ and $m_{\Omega^*_{bc}}(\frac{3}{2}^-) = 7.5~GeV$. 
 
According to the sum rules analysis, the working
regions of the continuum threshold $s_0$
and the Borel mass $M^2$ should be found by imposing the requirement that
the mass and residue exhibit good stability with respect to the variations
in these parameters. Therefore, we vary the continuum threshold $s_0$
and the Borel mass $M^2$, in order to find the ``working region"of $M^2$,
where the perturbative contribution is larger compared to the nonperturbative
part. Using the quark--hadron duality, the contributions of the  higher states and
continuum are taken as the perturbative ones starting from $s_0$.

The continuum threshold  depends on the energy in the vicinity of the
first excited state. In this respect, we choose the value of the continuum
threshold within the interval $s_0 = (100 - 125)~GeV^2$ for $bb$, $s_0 =
(50 - 65)~GeV^2$ for $bc$, and $s_0 = (14 - 22)~GeV^2$ for $cc$
baryons.

In the analysis of QCD sum rules, two conditions are satisfied for $M^2$.
a) The pole dominance with respect to the higher states and continuum; b)
The convergence of the OPE, i.e., dominance of the perturbative
part over the nonperturbative contributions.

The upper bound on $M^2$ can be obtained from the condition (a). For this
purpose, we introduce the ratio $R$, which describes relative contributions
of the continuum and pole,
\bea
\label{nolabel07}
R = {\ds \int_{s_0}^\infty ds \rho(s) e^{-s/M^2} \over \ds
\int_{(m_Q+m_Q^\prime)^2}^{\infty} ds \rho(s)e^{-s/M^2} }~.\nnb
\eea
Demanding that $R<1/2$, which guarantees that the pole contribution exceeds
the continuum and higher state contributions, we find the maximum values of
$M^2$ for $cc$, $bc$ and $bb$ baryons, as are listed below:
\bea
\label{nolabel08}
M_{max}^2 = \left\{ \begin{array}{c}
4.5~GeV^2~(\mbox{at}~\sqrt{s_0}=4.4~GeV),~\mbox{for}~\Xi_{cc}^\ast~\mbox{and}~\Omega_{cc}^\ast, \\
8.0~GeV^2~(\mbox{at}~\sqrt{s_0}=8.0~GeV),~\mbox{for}~\Xi_{bc}^\ast~\mbox{and}~\Omega_{bc}^\ast, \\
12.0~GeV^2~(\mbox{at}~\sqrt{s_0}=10.9~GeV),~\mbox{for}~\Xi_{bb}^\ast~\mbox{and}~\Omega_{bb}^\ast.
\end{array} \right.
\eea
The lower limit of $M^2$ can be obtained when the criteria (b) is satisfied.
Our numerical analysis leads to the following minimum values of $M^2$:
\bea                                                  
\label{nolabel09}
M_{min}^2 = \left\{ \begin{array}{c}
3.0~GeV^2~(\mbox{at}~\sqrt{s_0}=4.4~GeV),~\mbox{for}~\Xi_{cc}^\ast~\mbox{and}~\Omega_{cc}^\ast, \\
6.0~GeV^2~(\mbox{at}~\sqrt{s_0}=8.0~GeV),~\mbox{for}~\Xi_{bc}^\ast~\mbox{and}~\Omega_{bc}^\ast, \\
8.0~GeV^2~(\mbox{at}~\sqrt{s_0}=10.9~GeV),~\mbox{for}~\Xi_{bb}^\ast~\mbox{and}~\Omega_{bb}^\ast.
\end{array} \right.
\eea
In these ranges of the $M^2$, the relative contributions of the pole and continuum for each baryon are
presented in Table 2. In comparison, we also present the contributions of the pole and continuum obtained from the expressions presented in \cite{Rdhbtt10} in the same Table.
\begin{table}[h]
\renewcommand{\arraystretch}{1.3}
\addtolength{\arraycolsep}{1pt}
$$
\begin{array}{|c|c|c|c|c|}
\hline \hline
                & \mbox{Pole (Our Work)} & \mbox{Continuum (Our Work)} &\mbox{Pole\cite{Rdhbtt10}}&\mbox{Continuum\cite{Rdhbtt10}}\\ \hline
\Xi_{cc}^\ast     & (63-71)\% & (29-37)\%  &57\%& 43 \%        \\
\Omega_{cc}^\ast  & (75-81)\% & (19-25)\%  &{\mbox{---}}& {\mbox{---}}       \\
\Xi_{bc}^\ast     & (68-75)\% & (25-32)\%  &{\mbox{---}}& {\mbox{---}}       \\
\Omega_{bc}^\ast  & (77-83)\% & (17-23)\%   &{\mbox{---}}&  {\mbox{---}}     \\
\Xi_{bb}^\ast     & (52-58)\% & (42-48)\%   &57\%& 43 \%     \\ 
\Omega_{bb}^\ast  & (66-70)\% & (30-34)\%   &{\mbox{---}}&   {\mbox{---}}    \\
\hline \hline
\end{array}
$$
\caption{The relative contributions of the pole and continuum to the sum
rule in respect to the variation of $M^2$ in the ``working region", together with those obtained  from the expressions presented in \cite{Rdhbtt10}.}
\renewcommand{\arraystretch}{1}
\addtolength{\arraycolsep}{-1.0pt}
\end{table}

Using the the working regions of $M^2$
and $s_0$, we obtain the results for
the masses of spin--3/2 doubly heavy baryons, which are all presented in
Table 3. For completeness,  we present the results of the other works in the same
Table as well. The residues of these baryons are also presented in Table 4.
We see from these Tables that our results on the masses are overall very close  to the
values given in \cite{Rdhbtt06},
\cite{Rdhbtt10}, \cite{Rdhbtt11}, \cite{Rdhbtt14} and \cite{Rdhbtt15}.
For the masses of $\Xi_{bb}^\ast$ and
$\Omega_{bb}^\ast$, the predictions of all approaches are very close to each
other. On the other hand, for the masses of $\Xi_{bc}^\ast$ and
$\Omega_{bc}^\ast$, the predictions of \cite{Rdhbtt11} are slightly larger in magnitude
while the results of \cite{Rdhbtt06} are slightly  smaller 
compared to our results for central values.
As far as the residues of the $\Xi_{cc}^\ast$ and $\Omega_{cc}^\ast$ baryons are
concerned, our predictions are higher in magnitude compared to the ones
presented in \cite{Rdhbtt10} and \cite{Rdhbtt14}. Our prediction for the residue of the $\Xi_{bb}^\ast$ baryon is also higher when compared to that of the \cite{Rdhbtt10}, 
 while our predictions on the residues of the $\Xi_{bb}^\ast$ and
$\Omega_{bb}^\ast$ baryons almost match with those of the \cite{Rdhbtt14}.

\begin{table}[h]

\renewcommand{\arraystretch}{1.3}
\addtolength{\arraycolsep}{-0.5pt}
\small
$$
\begin{array}{|l|r@{}l|r@{}l||r@{}l|r@{}l|r@{}l|r@{}l|}
\hline \hline  
 \multirow{2}{*}{}                                                                &
 \multicolumn{4}{c||}{\multirow{1}{*}{\mbox{Our Work}}}                           &  
 \multicolumn{2}{c|}{\multirow{2}{*}{\mbox{\cite{Rdhbtt10} and \cite{Rdhbtt15}}}} &
 \multicolumn{2}{c|}{\multirow{2}{*}{\mbox{\cite{Rdhbtt11}}}}                     &
 \multicolumn{2}{c|}{\multirow{2}{*}{\mbox{\cite{Rdhbtt06}}}}                     &
 \multicolumn{2}{c|}{\multirow{2}{*}{\mbox{\cite{Rdhbtt14}}}}                       \\
	                                                                          &
 \multicolumn{2}{c}{\multirow{1}{*}{\mbox{~Structure $\rlap/{q}g_{\mu\nu}$~}}}    & 
 \multicolumn{2}{c||}{\multirow{1}{*}{\mbox{~Structure $g_{\mu\nu}$~}}}           &
 \multicolumn{2}{c|}{}                                                            &
 \multicolumn{2}{c|}{}                                                            &
 \multicolumn{2}{c|}{}                                                            &
 \multicolumn{2}{c|}{}                                                              \\ \hline 
 \Xi_{cc}^\ast     &~~~~3.69\pm&0.16   &~~~3.72\pm&0.18  &     3.58\pm& 0.05 &  3.90\pm&0.10   &3.&727  &     3.61\pm&0.18 \\
 \Omega_{cc}^\ast  &    3.78\pm&0.16   &   3.78\pm&0.16  &     3.67\pm& 0.05 &  3.81\pm&0.06   &3.&872  &     3.76\pm&0.17 \\ 
 \Xi_{bb}^\ast     &    10.4\pm&1.0    &  10.3\pm&0.2    &    10.33\pm& 1.09 & 10.35\pm&0.08   &10.&237 &    10.22\pm&0.15 \\
 \Xi_{bc}^\ast     &    7.25\pm&0.20    &   7.2\pm&0.2    &{\mbox{---}}&      &  8.00\pm&0.26   &6.&98   &{\mbox{---}}&     \\
 \Omega_{bc}^\ast  &    7.3\pm&0.2     &   7.35\pm&0.25   &{\mbox{---}}&      &  7.54\pm&0.08   &7.&13   &{\mbox{---}}&     \\
 \Omega_{bb}^\ast  &   10.5\pm&0.2     &  10.4\pm&0.2    & 10.38\pm& 1.10    & 10.28\pm&0.05   &10.&389 &    10.38\pm&0.14 \\
 \hline \hline
\end{array}
$$

\caption{The mass spectra of the spin--3/2 doubly heavy baryons 
in units of $GeV$.}

\renewcommand{\arraystretch}{1}
\addtolength{\arraycolsep}{-1.0pt}

\end{table}
\begin{table}[h]

\renewcommand{\arraystretch}{1.3}
\addtolength{\arraycolsep}{-0.5pt}
\small
$$
\begin{array}{|l|r@{}l|r@{}l||c|r@{}l|}
\hline \hline  
 \multirow{2}{*}{}                                            
 &\multicolumn{4}{c||}{\multirow{1}{*}{\mbox{Our Work}}}      &   
 \multicolumn{1}{c|}{\multirow{2}{*}{\mbox{\cite{Rdhbtt10}}}} &
 \multicolumn{2}{c|}{\multirow{2}{*}{\mbox{\cite{Rdhbtt14}}}}   \\
	                                                      &
 \multicolumn{2}{c}{\mbox{~Structure $\rlap/{q}g_{\mu\nu}$~}} & 
 \multicolumn{2}{c||}{\mbox{~Structure $g_{\mu\nu}$~}}        &
 \multicolumn{1}{c|}{}                                        &
 \multicolumn{2}{c|}{}                                          \\ \hline 
 \Xi_{cc}^\ast     &~~~~0.12\pm&0.01    &~~~0.12\pm&0.01    &   0.071\pm0.017  &     0.070\pm&0.017  \\
 \Omega_{cc}^\ast  &    0.14\pm&0.02    &   0.13\pm&0.01    &        {\mbox{---}}                    &     0.085\pm&0.019  \\ 
 \Xi_{bb}^\ast     &    0.22\pm&0.03    &   0.21\pm&0.01    &   0.111\pm0.040 &     0.161\pm&0.041  \\
 \Xi_{bc}^\ast     &    0.15\pm&0.01    &   0.15\pm&0.01    &        {\mbox{---}}                    &     {\mbox{---}}&  \\
 \Omega_{bc}^\ast  &    0.18\pm&0.02    &   0.17\pm&0.01    &        {\mbox{---}}                    & {\mbox{---}}&       \\
 \Omega_{bb}^\ast  &    0.25\pm&0.03    &   0.25\pm&0.02    &        {\mbox{---}}                    & 0.199\pm&0.048      \\
 \hline \hline
\end{array}
$$

\caption{The residues of the spin--3/2 doubly heavy baryons in units of $GeV^3$.}

\renewcommand{\arraystretch}{1}
\addtolength{\arraycolsep}{-1.0pt}

\end{table}

As the final remark, we note that in the case we neglect or take into account
the contributions coming from the negative parity  baryons  the results change
less than  $5\%$. 

\section{Conclusion}

In the present work we  calculated  the masses and residues of the doubly heavy  spin--3/2 baryons
 within the QCD sum rules method. In calculations we  took into account the contributions of the operators up to five dimensions in OPE. We also included the contributions of the negative parity baryons. 
We compared our predictions on the masses and residues with the existing predictions in the literature. Our results on the masses of the doubly heavy  spin--3/2 baryons are overall consistent with the previous predictions
of different works discussed in the body text. However in the case of residues, although our results for some baryons  are in good consistency with the results of some works,  our predictions for some other  baryons deviate considerably from the existing predictions in the literature.
 We expect that our  predictions on the masses and residues in this manuscript  can all be checked at LHCb in near future.

\end{document}